\documentclass[11pt,preprint]{aastex}
\usepackage{psfig}

\newcommand\nudotdotdot{\ifmmode\stackrel{\bf \,...}{\textstyle
\nu}\else$\stackrel{\,...}{\textstyle \nu}$\fi}

\newcommand{\approxgt}{\mbox{$\;^{>}\hspace{-0.24cm}_{\sim}\;$}}

\shorttitle{A Low Cost Distributed Computing Approach to Pulsar
Searches at a Small College}
\shortauthors{Cantino et al.}

\begin{document}

\title{A Low Cost Distributed Computing Approach to Pulsar Searches at
a Small College\footnote{Adapted from a talk given at the Eleventh
SIAM Conference on Parallel Computing, February 25-27, 2004 in San
Francisco, CA.}}

\author{Andrew Cantino, 
Fronefield Crawford, 
Saurav Dhital,
John P. Dougherty, 
Reid Sherman}
\affil{Haverford College, 370
Lancaster Ave., Haverford, PA 19041}

%\email{fcrawfor@haverford.edu}

\begin{abstract}
We describe a distributed processing cluster of inexpensive Linux
machines developed jointly by the Astronomy and Computer Science
departments at Haverford College which has been successfully used to
search a large volume of data from a recent radio pulsar survey.
Analysis of radio pulsar surveys requires significant computational
resources to handle the demanding data storage and processing needs.
One goal of this project was to explore issues encountered when
processing a large amount of pulsar survey data with limited
computational resources.  This cluster, which was developed and
activated in only a few weeks by supervised undergraduate summer
research students, used existing decommissioned computers, the campus
network, and a script-based, client-oriented, self-scheduled data
distribution approach to process the data. This setup provided
simplicity, efficiency, and ``on-the-fly'' scalability at low
cost. The entire 570 GB data set from the pulsar survey was processed
at Haverford over the course of a ten-week summer period using this
cluster. We conclude that this cluster can serve as a useful
computational model in cases where data processing must be carried out
on a limited budget. We have also constructed a DVD archive of the raw
survey data in order to investigate the feasibility of using DVD as an
inexpensive and easily accessible raw data storage format for pulsar
surveys. DVD-based storage has not been widely explored in the pulsar
community, but it has several advantages. The DVD archive we have
constructed is reliable, portable, inexpensive, and can be easily read
by any standard modern machine.
\end{abstract}


\section{Introduction and Motivation}

Radio pulsars, which are exotic, rapidly spinning collapsed stars, are
named for the characteristic radio pulses that make them detectable.
Pulsars are not completely understood, but it is thought that they
form when a massive star explodes and the core collapses into a
super-dense star called a neutron star.  Extreme electromagnetic
fields accelerate charges, and a beam of radio waves is emitted, which
is seen as a pulsed source as the pulsar spins. Pulsars were first
detected accidentally around 35 years ago \citep{hbp+68} and they have
since been found with spin periods as fast as 1.56 ms \citep{bkh+82}.

Pulsars are interesting from both astronomical and physical
standpoints.  Discovering more pulsars aids in the understanding of
the underlying Galactic pulsar distribution and population as well as
pulsar formation and evolution.  Pulsars can also be used as probes of
the interstellar medium by observing the effects of radio wave
propagation through interstellar plasma. Pulsars are also useful for
testing physical theories, such as General Relativity (e.g., Taylor
1994\nocite{t94}), and the physics of the pulsar emission process.
Pulsars provide a rare opportunity to study extreme physics with their
tremendous magnetic fields and huge densities.  Their extremely
regular pulse periods and unusual physical conditions make pulsars
excellent tools for observational physics. Complete reviews of
pulsars, their properties, and their uses are presented in
\citet{mt77} and \citet{lg98}.

In order to find and study pulsars, very large data sets must be
searched for their characteristic periodicities. This analysis
requires significant computational resources in the form of processing
power and data storage capability and data accessibility.  At a small
college with a limited research budget, these resources may not be
readily available, making pulsar data processing difficult or
impossible unless alternative options can be explored.  The Astronomy
and Computer Science departments at Haverford College have jointly
developed an inexpensive distributed processing cluster of
decommissioned machines running Linux.  We aimed to explore the issues
encountered when searching for pulsars with old, decommissioned
computers in a networked cluster. Our goal was not only to discover
new pulsars in the survey, but also to test the feasibility of using
such a cluster for large-scale pulsar survey processing in the small
college research environment.

The survey, which was conducted in collaboration with the pulsar group
at McGill University, targeted 56 unidentified gamma-ray sources from
the 3rd EGRET catalog \citep{hbb+99} at intermediate Galactic
latitudes. The survey used the multibeam receiver \citep{swb+96} on
the Parkes 64-meter radio telescope in Parkes, Australia. This
combination of instruments has been very successfully used to find
pulsars in previous radio pulsar surveys \citep{mlc+01, mhl+02,
kbm+03, ckm+00, ebs+01}. It is possible that some of the target
gamma-ray sources could be powered by previously unidentified
energetic radio pulsars. Radio pulsar counterparts to these sources
would not only be interesting systems to study individually (e.g.,
Roberts et al. 2002\nocite{rhr+02}), but they would also help resolve
outstanding questions about the pulsar emission mechanism and the
physical origin of pulsar radiation at different wavelengths (see,
e.g., Harding et al. 2004 and references therein\nocite{hgg+04}).
More detailed descriptions of this survey, including the science
goals, results, and implications of the survey, are presented
elsewhere \citep{rlh+03, rrh+03}.

We discuss the development, deployment, advantages, and disadvantages
of our chosen method of parallelization of the computational problem
at hand.  Our data processing problem, though quite large, is
embarrassingly parallel \citep{wa99}, meaning that the task can be
easily divided up and tasked to separate client machines for local
processing without additional interprocess communication.  After
consideration of various methods of implementation, we settled on a
simple setup that used existing low-cost technologies.  The benefits
of this approach included ease of adding additional processors,
modular design, low cost, and rapid deployment. This set of benefits
was well-suited to our small college research program and can serve as
a computational model for similar institutions.

We have also constructed an archive of the raw survey data on DVD in
order to investigate the feasibility of using DVD as an inexpensive
and accessible format for raw data storage for pulsar
surveys. DVD-based storage has not been widely explored in the pulsar
community for data storage. We describe this archive and the
advantages of using DVD for pulsar survey data storage. We also
compare the costs of using DVD vs.~Digital Linear Tape IV (DLT-IV), a
commonly used data storage format in pulsar astronomy.

\section{Survey Data and Data Processing Details}

Each of the 56 unidentified EGRET gamma-ray sources targeted in the
survey was observed with the 1400-MHz 13-beam multibeam receiver on
the Parkes radio telescope. Owing to the uncertainty in the position
of each EGRET source (which typically has an uncertainty of about 1
square degree), full spatial coverage of each source required four
multibeam pointings of 13 beams each (see, e.g., Staveley-Smith et
al.~1996\nocite{swb+96}; Manchester et al.~2001\nocite{mlc+01}). A
total of 3016 beams were recorded at the telescope to be processed in
the survey\footnote{Nine pointings were observed twice, and one
pointing was not observed.  All other pointings were unique.}.  For
each telescope pointing, we used a 35-minute observation time which
was sampled at 0.125 ms with 1-bit per sample.  96 contiguous
frequency channels of 3 MHz each were recorded, providing a total
observing bandwidth of 288 MHz. The observing setup was similar to
that described in detail by \citet{mlc+01} for the Parkes Multibeam
Pulsar Survey. Each resulting beam contained 194.8 MB of raw data,
corresponding to a total of 573.7 GB of raw survey data to be stored
and processed.

When a radio pulse from a distant pulsar travels through space, plasma
in the interstellar medium disperses the pulse, causing lower
frequency components to be delayed relative to higher frequency
components.  This dispersion effect, quantified by a dispersion
measure (DM), is one indicator of the distance to the pulsar.  The DM
is the integrated column density of free electrons between Earth and
the pulsar (e.g., Manchester \& Taylor 1977\nocite{mt77}).  To detect
a pulsar, one must first correct for dispersion by appropriately phase
adjusting each frequency channel. However, if one does not know {\it a
priori} how far away the pulsar is, or if a pulsar signature even
exists in the data, the DM to correct for is unknown.  The data must
therefore be dedispersed at a large number of trial DMs, with each
trial separately searched for a pulsar signature.  In general, aside
from an initial processing overhead, searching in dispersion measure
space increases the complexity of the processing linearly by a
function of the number of trial dispersion measures. For our search,
we dedispersed each data set at 450 trial DMs, ranging from 0 to 700
pc cm$^{-3}$. This easily encompassed the expected maximum DM for
Galactic pulsars in the directions observed \citep{cl02}.\footnote{One
pc, which stands for parsec, is equal to $3.0856 \times 10^{18}$ cm.}

After each trial dedispersion, the frequency channels were summed to
create a time series. This time series was then Fourier transformed
via a Fast Fourier Transform (FFT) to identify strong signal
candidates in the resulting power spectrum.  The characteristic signal
of interest is a wideband, extremely regular series of radio pulses.
FFTs are well equipped for finding this type of signal, although radio
frequency interference (RFI) can mask these signatures.  To mitigate
RFI, we filtered our data for certain known interference signals.

The raw data from the survey were originally processed at McGill
University using custom pulsar search software \citep{r01,
rem02}. During this first processing run, several new pulsars were
discovered, and all previously known pulsars that were spatially
coincident with the target sources were redetected. The second pass at
processing, which is described here, was conducted at Haverford
College and used different software \citep{lkm+00}. The analysis at
Haverford was primarily done by undergraduates during a ten-week
summer research period.  The reprocessing of the data aimed to see
whether there were pulsars that were missed during the first
processing pass. Of particular interest were long period pulsars ($P
\approxgt 20$ ms); fewer than expected were found in the first
processing run.  We therefore aggregated the data prior to processing
to reduce the data size and thus significantly decrease the processing
time while still maintaining sensitivity to longer-period pulsars.
The data were aggregated into contiguous groups of 4 frequency
channels (effectively reducing the resolution from 3 MHz to 12 MHz)
and into contiguous groups of 16 time samples (effectively reducing
the sampling time from 0.125 ms to 2.0 ms). This reduced the size of
each data set by a factor of 64, which greatly eased the computational
burden.

\begin{figure}
\centerline{\psfig{figure=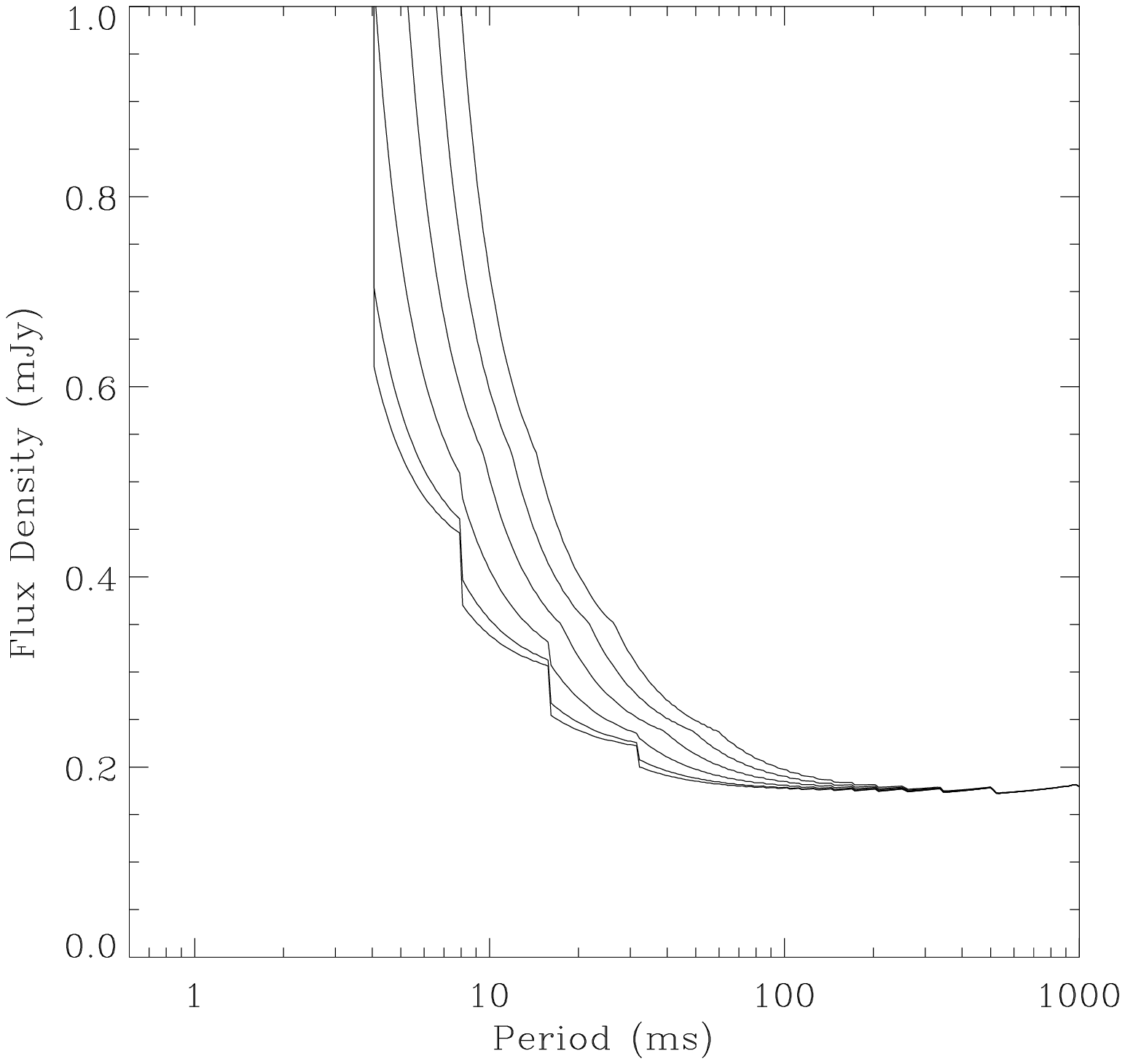,width=7in}}
\caption{Minimum detectable 1400 MHz flux density as a function of
pulsar period for the EGRET survey described here. The six curves
correspond to assumed DM values of 0, 20, 40, 60, 80, and 100 pc
cm$^{-3}$ (in order from left to right). An intrinsic duty cycle of
5\% for the pulsed emission is assumed in the sensitivity
calculation. For pulsar periods faster than about 20 ms, the
sensitivity to pulsations is sharply degraded. The details of the
sensitivity calculation are outlined in \citet{c00} and
\citet{mlc+01}. This plot applies to the aggregated (resampled) data
processed at Haverford College using the computer cluster described
here.}
\label{fig-2}
\end{figure}

According to the Nyquist theorem \citep{n28}, we were in principle
sensitive to pulsars with periods as fast as 4 ms in the aggregated
data, but in practice the sensitivity was degraded for periods below
about 20 ms (this is partially due to intra-channel dispersion effects
which cannot be corrected for in the processing).  This sensitivity
degradation is clearly shown in Figure \ref{fig-2}, which shows
minimum detectable 1400 MHz flux density as a function of pulsar
period for a range of assumed pulsar DMs.  We redetected all of the
new and previously known pulsars which had been found in the first
processing run in this period range, thereby demonstrating the
effectiveness of this approach.

\section{Possible Computing Approaches}

In order to practically process the large amount of data from the
survey, we needed to use multiple processors in a cluster.  There were
several possible approaches to parallelizing the processing, but the
choices were greatly limited by the technology and facilities locally
available to us. For the processing at Haverford College, we had seven
older 400-MHz Intel Celeron PCs, one new 2.5-GHz Pentium 4 PC, and the
Haverford College network at our disposal.  We ran Red Hat 9.0 Linux
on these machines.  We also had access to a number of existing
software packages designed for pulsar searching.

Given the available technology, our choice was between implementing a
parallel program using a message-passing library, such as MPI/PVM, or
writing our own distribution and processing system.  While using MPI
could enhance performance, it required either writing new MPI-enhanced
parallel software or extensive porting of existing software.  In
either case, this would have been a time consuming task which would
have delayed processing in the limited time available.  We found that
the existing software would compile and run on Red Hat 9.0 and could
do all of the needed processing. We decided to implement our own
distribution system, which is described in the next section.

\section{Computer Cluster Details}

The initial pulsar search cluster consisted of a set of 8 PCs running
Linux (Red Hat 9.0) and two 180 GB disks. One of the PCs (a 400-MHz
Intel Celeron with 128 MB RAM) was a dedicated file server and
workstation, but did no actual data processing.  Another machine (a
2.5-GHz Pentium 4 with 512 MB RAM) served simultaneously as a
secondary file server (data only) and processing client.  The
remaining six 400-MHz Celerons, with 128 MB RAM each, were dedicated
processing clients.  In terms of cluster nomenclature, the cluster was
not technically a Beowulf because it lacked a dedicated network, and
some machines performed multiple user-based tasks.  Rather, the
cluster can be best described as a semi-dedicated Network of
Workstations (NOW) \citep{cs99} running a data-parallel type algorithm
taking advantage of the existing data partitioned 195 MB unaggregated
beam structure \citep{ggk+03}.

The cluster was divided into clients and servers, with all control
left in the hands of the clients, allowing for self-maintained load
balancing.  The servers used Network File System (NFS) file servers
that were remotely mounted by every client machine.  The NFS server
hosted a flat database file that contains a list of beams, their
current processing status, and location information.  It also stored
the raw data file for each beam and the results returned by the
clients.

Client machines ran scripts written in PERL that parsed the database
on the server, found available beams for processing, downloaded them
to their local hard drives, and ran a series of pre-existing commands
on the local data.  Each client processed one beam at a time.  This
resulted in a nearly linear increase in processing potential as more
clients were added, limited only by network load (discussed below).
Once a client completed processing, it returned the relevant files to
the server and looked for more work to do.

System status and control were achieved through control files on the
server.  Clients maintained individual log files, status files, and
timing files on the server.  They also checked certain server-based
text files for instructions.  This method of control allowed for
monitoring and maintenance of the cluster, either through specialized
monitoring scripts or a text editor.  In this state, the clients
allowed for limited error checking and recovery, with the ability to
recover from bad data files by marking them as failed and moving on to
new data.

\begin{figure}
\epsscale{1.0}
\vspace{9.5in} 
\hspace{-2.0in} 
\plotone{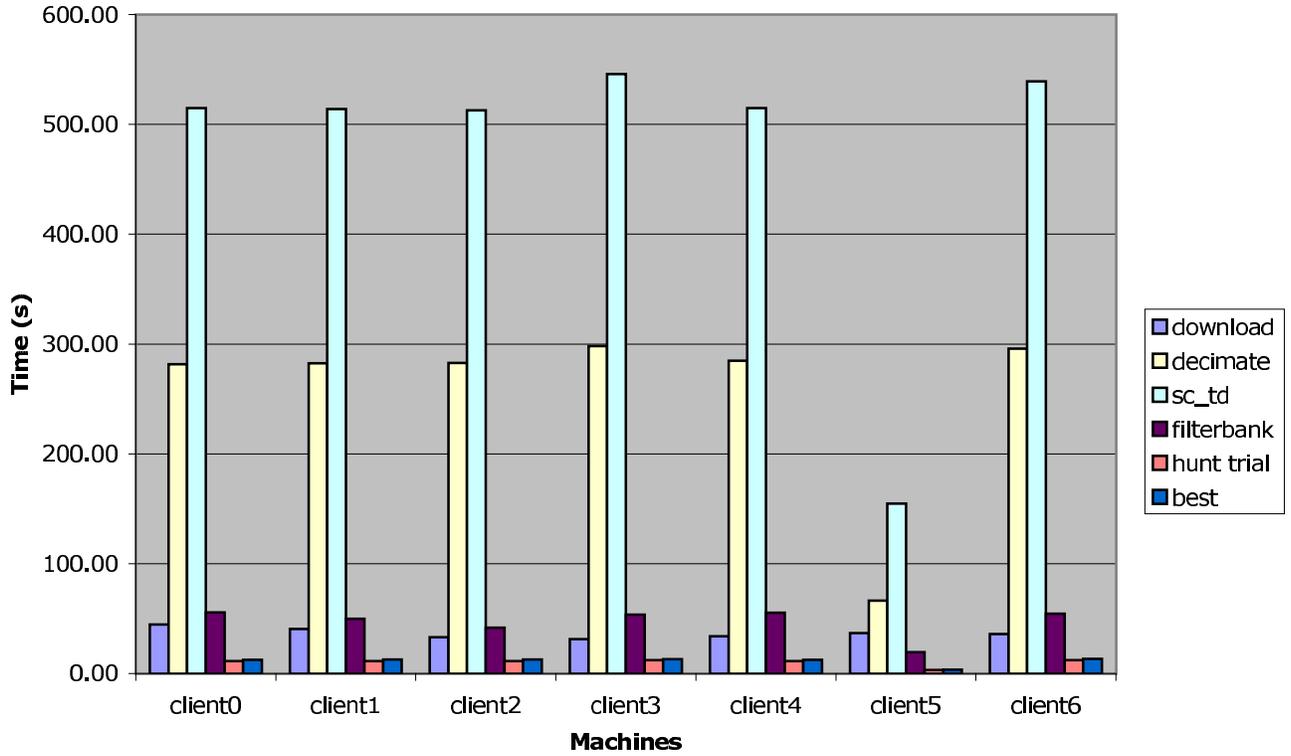} 
\vspace{-3.0in} \figcaption[graph4aa.ps]{Timing statistics for the
seven client machines in the initial processing cluster. The average
times for function completion are shown for six 400-MHz Intel Celerons
and one 2.5-GHz Pentium 4 (client5). {\it Download} is the initial
data transfer process, {\it decimate} is the data aggregation process,
{\it sc\_td} and {\it filterbank} put the raw data into a readable
format, {\it hunt trial} runs through a single DM trial and computes a
$2^{20}$ point FFT on the dedispersed time series (this step was
repeated 450 times in the processing using different trial DMs), and
{\it best} picks out the most promising pulsar candidates at the end
of the processing. The total processing time for client5 was faster by
about a factor of three than the other machines.  Table \ref{tbl-1}
lists additional cluster timing results.  \label{fig-3}}
\end{figure}

Performing the full analysis (all 450 DM trials) on a single
aggregated beam with the modern 2.5-GHz Pentium 4 computer took on
average about 28 minutes.  On one of the older computers (400-MHz
Intel Celeron), it took on average approximately 100 minutes (only
about a factor of three longer).  Of this time, approximately 36
seconds on average was spent downloading the data to the local disk
(corresponding to a 5.4 MB/s transfer rate over the local
network). This download time was roughly constant for all machines
(independent of the processor speed) since the download was limited by
the network transfer speed.  All other functions in the processing
were slower by about a factor of three for the slower machines
compared to the modern computer.  Figure \ref{fig-3} and Table
\ref{tbl-1} give more detailed timing information for the cluster.

\section{Benefits of Our Cluster Processing Approach}

There were several significant benefits in using this cluster
processing approach.

\begin{enumerate}

\item {\bf Scalability.} Because our cluster was completely maintained
by the clients themselves, it was very simple to add an additional
client without the need to restart the cluster.  One simply remote
mounted the NFS drive on a Red Hat 9.0 machine, copied two files to
the local drive, created two files and a directory, and edited one
configuration line.  All other needed files were loaded from the
server as needed.  With practice, it took less than 5 minutes to get
an additional (preinstalled) Red Hat 9.0 machine up and added to the
cluster. Additional machines have since been added to the cluster as
they have been decommissioned in order to increase the processing
capability of the cluster for future projects.

\item {\bf Use of Existing Technology.} Instead of having to rewrite
signal-processing software for use with MPI, we were able to use
existing software written in both C and Fortran. Specifically, we used
the pulsar search packages Seek and Sigproc written by Dunc Lorimer
(e.g., Lorimer et al 2000\nocite{lkm+00})\footnote{See also
http://www.jb.man.ac.uk/$\sim$drl.}.  Additionally, we used PERL, NFS,
and the Linux operating system, all of which are well-tested and
freely available products.

\item {\bf Simplicity.} For communication, we used plain text files
that were edited with either custom scripts or a text editor.  This
allowed for quick debugging and error detection, as well as convenient
control of the cluster remotely over a secure shell.  Additionally,
since all the scripts were plain text files, making changes simply
required editing the script and restarting the client processes.

\item {\bf Efficiency.} Our use of existing technology allowed for
rapid development and deployment.  The cluster was fully operational
and searching for pulsars in a few weeks.  This meant that the survey
processing could be completed in the ten-week summer research period.

\item {\bf Near-linear Speedup.}  We found that, on average, the older
400-MHz Celerons took about 100 minutes to fully process a single
beam, while the new 2.5-GHz Pentium 4 took only 28 minutes.  Of this
processing time, both systems averaged a network download time of just
36 seconds, comprising six tenths of a percent of the total processing
time for the older systems and two percent for the newer machine (see
Table \ref{tbl-1} for additional timing results).  Because the
download-to-computation time ratio was so small, we would expect a
near-linear increase in processing capability as additional computers
are added to the cluster.  This linear increase should hold until
download times start to overlap, at which point processing times would
slowly increase.  However, the cluster software used dedicated file
downloads with scheduling scripts with built-in download staggering to
avoid overlapped downloads.

\item {\bf Dependability.}  The cluster was maintained using detailed
client logs, timing records and monitoring scripts. Errors from
incompletely processed data sets were recorded for later reprocessing.
The existence of many large data sets, each implying substantial
computation time per processing node, motivated the use of
self-scheduling \citep{hsf92} at the beam level.  Benefits of this
approach included ease of implementation, near maximal throughput, and
trackable progress.

\item {\bf Cost.} The costs of developing the cluster were kept low,
since Red Hat 9.0 (Linux), PERL, and NFS are all free products, as are
Seek and Sigproc.  Additionally, we were able to use older
decommissioned computers, readily available from computer labs and
other sources.  A cursory analysis of prices on eBay in July 2003
showed bulk lots of older computers available. For example, fifty
350-MHz Pentium II machines were selling for for \$3500 (\$70 per
machine), and twenty-five 500-MHz Pentium III machines were selling
for \$3000 (\$120 per machine).  Either of these lots would offer
significantly more raw processing power than the equivalent cost in
new machines for our purpose.  Our cluster cost us \$360 for two 180
GB hard drives and \$835 for the new Pentium 4 machine.  The older
Celerons were scavenged, and we used the available campus network for
networking.  The approximate cost of the entire cluster was \$1200,
not including the additional hidden costs of using the college
network, lab space, and power. Student stipend costs are also not
included in this figure. Developing the cluster at low cost was an
important consideration since research funds were limited.

\end{enumerate}

\section{Drawbacks of Our Cluster Processing  Approach}

The problems we encountered with our cluster were primarily easily
fixed coding flaws or manageable technical issues.  The largest
problem was also one of our benefits: the lack of a central management
server.  While this provided for excellent modularity and ease of
development, it also made the cluster harder to maintain and check for
errors during processing.  Still, we found that the trade off for
rapid development was worthwhile.  We were able to start the data
processing in a couple of weeks because we did not need to write
server or signal processing software.
                                                                               
Other issues we encountered included troublesome remote file locking
(we settled this by using three separate file locking mechanisms
simultaneously) and coding issues.  Additionally, network security was
a concern because we used the campus network for client-server
communication instead of a dedicated network.  This required that we
keep all computers up-to-date and secure.  This is an intrinsic
problem with NOW-type architectures \citep{cs99}.

Our approach worked very well for an embarrassingly parallel data
processing and distribution problem with large data sets (such as the
pulsar search described here), but it would be inappropriate for
problems requiring interprocess communication.  For such problems, a
language with a richer set of communication primitives, such as MPI,
would be more appropriate.

\section{Motivation for a DVD Archive of the Raw Survey Data} 

We have also constructed a DVD archive of the raw survey data that
were processed with the cluster in order to investigate the
feasibility of using DVD as an inexpensive and easily accessible
format for raw data storage for pulsar surveys. The DVD format has not
yet been widely explored in the pulsar community, but it has several
advantages. The archive we have made is reliable, portable,
inexpensive, and easily read on any standard modern machine with a
DVD-ROM drive.  The archive is presently available for extraction of
specific data from the survey if needed for reprocessing at any time
in the future.

The raw survey data were originally recorded and stored on DLT-IV tape
at the Parkes telescope. DLT-IV is the standard medium used for pulsar
data recording at several observatories (including Parkes).  DLT
systems are used because they can record vast amounts of data in real
time at the telescope.  DLT-IV tape media can hold up to 35 GB of
uncompressed raw data per tape.  The main motivation for using DLT,
aside from its real time data acquisition capability, is convenience
in data consolidation, since a relatively small number of DLT tapes
can often hold all the data for a given pulsar survey project.
However, this particular convenience comes at a significant cost.  DLT
8000 tape drives, which are required to read the tapes, sell for
approximately \$2300\footnote{All prices quoted in this paper are from
June 2003 unless otherwise specified.}, a sizable investment for
research programs at small colleges which have very limited research
budgets.  These drives are required not only for writing the data (for
instance, at the telescope) and for making backups of the data, but
also for reading the data from tape before data processing can even
begin. Furthermore, the DLT-IV storage medium itself costs about \$65
per tape.  Since the medium is magnetic and sequentially organized
(i.e., implying a single point of failure), the tapes can be subject
to wear and catastrophic failure through repeated usage as well as
from unregulated environmental conditions (e.g., heat and
humidity). With a single point of failure, an entire tape is lost. In
some cases, this can correspond to entire days of telescope time being
lost.  The failure rate of DLT-IV tapes in the Parkes Multibeam Pulsar
Survey \citep{mlc+01, mhl+02, kbm+03} was a few percent, which, for a
large pulsar survey, adds significantly to the expense of the project
when re-observation costs are factored in. At a small college such as
Haverford, the usefulness of a DLT drive would also be limited since
its use is so specialized. A DVD writer/reader is more generically
useful for other research projects in departments across the college,
particularly if the drive is external and can be easily moved to
different machines and locations.

Since DVD drives are now available on standard modern computers, a DVD
archive can be easily accessed for reprocessing at a later time.  No
specialized drive is needed, making the data very portable across many
platforms (Linux, Windows, Mac) and accessible to researchers who do
not have the funds for specialized peripherals, such as DLT
drives. The DVDs themselves are also easily transported and can be
copied and shipped in the mail with little cost or worry about
failure.  DVDs are also not as susceptible to changes in environmental
conditions which affect reliability. For instance, no environmentally
controlled room is required for storage, and DVDs are not subject to
wear through contact during many tape passes.

DVD writers can be purchased for about \$400, which is very affordable
for observatories or research groups that are charged with
maintaining, copying, or disseminating raw pulsar data. DVD media are
also inexpensive and can be obtained for a cost of less than \$1.50
per disc. Table \ref{tbl-2} outlines the cost structure for storing
the EGRET survey raw data on DVD vs.~DLT-IV tape. Archiving the data
on DVD is significantly less expensive.

\section{DVD Archive Details}

\begin{figure}
\vspace{1in}
\centerline{\psfig{figure=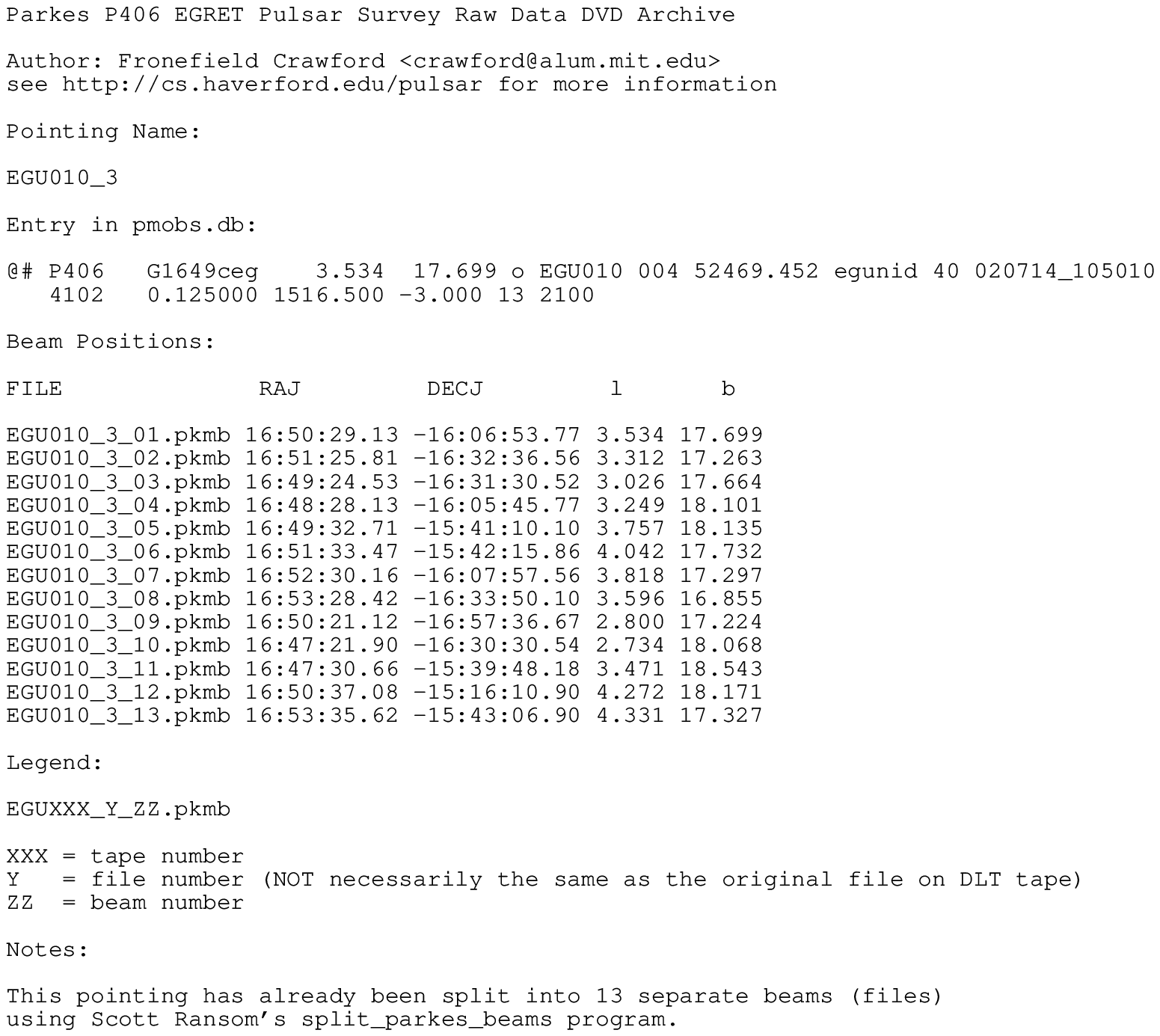,width=7in}}
\vspace{-3in}
\caption{Example plain text description file included on each DVD
archive disc.  This file, which was automatically generated using a
custom PERL script, describes the details of the particular pointing
observation archived on the DVD.  Each disc contains a single multibeam
pointing (13 beams), corresponding to about 2.5 GB of storage space used 
on the disc.}
\label{fig-4}
\end{figure}

The complete EGRET survey consisted of 3016 distinct beams of 194.8 MB
each, totaling 573.7 GB of raw data to be archived on DVD. In
constructing the DVD archive, each pointing was remotely read off DLT
tape from a drive located at McGill University, split into 13 separate
beams, then transferred over the internet to a local machine in our
cluster at Haverford. The 13 split beams of raw data from a single
pointing were then written to a single DVD along with a plain text
file describing the contents of the DVD (including a listing of the 13
beam positions).  An example of one of these text files is presented
in Figure \ref{fig-4}.  Each disc in the DVD archive contains about
2.5 GB of data, corresponding to an efficiency of use of the storage
space of about 53\%. This made the archive more manageable than it
would be if we had split pointings across discs.  A complete index of
the DVD archive is accessible via the world wide
web\footnote{http://cs.haverford.edu/pulsar}. Information about each
EGRET target source is also presented on this page, as is observing
information about each individual pointing. The individual beam
position (in both J2000 celestial coordinates and Galactic
coordinates) for every beam in the survey is also listed on this web
page, making identification of the relevant DVDs and extraction of the
data for reprocessing easy.

In making the DVD archive, we compiled a comparison of the costs of
using DVD vs. DLT-IV tape for storing our survey data (Table
\ref{tbl-2}).  Even when using the DVD storage space inefficiently for
the sake of ease of data management (53\% of space used vs. 75\% for
the 22 DLT-IV tapes in the EGRET survey), the cost per GB of storage
for DVD is still better by a factor of about four (\$0.55/GB
vs. \$2.48/GB for DLT-IV). The figure improves to a factor of six in
favor of DVD if both DVD and DLT storage space are used 100\%
efficiently. If one includes the auxiliary costs of creating and
maintaining the survey in both cases, then the total cost of creating
the archive (including a DVD writer, DVD carrying case, and a DLT 8000
drive) would be cheaper by a factor of five in the case of DVD (\$760
for DVD vs. \$3730 for DLT)\footnote{This does not include the cost of
student labor to burn the DVDs for the archive.}.

DVD is the more reliable, portable, and inexpensive storage medium for
long-term archiving of raw pulsar survey data. These benefits are
offset only by the relatively small storage capability of each
individual DVD (currently a maximum of 4.7 GB), which is an important
issue if individual observations produce data files in excess of this
limit. However, the small storage capability of DVD also has an
advantage in terms of the amount of data lost in the event of failure.
A DVD writer could be purchased for about \$400 and installed at any
observatory for immediate backup of pulsar data and for easy
dissemination (copying and mailing) of archived data to researchers
who are off-site.

\section{Results, Conclusions, and Possible Future Work}

All 3016 beams from the radio pulsar survey of 56 unidentified EGRET
sources, corresponding to 573.7 GB of raw survey data, were processed
with the networked Linux cluster developed by the Astronomy and
Computer Science departments at Haverford College.  Most of this work
was conducted by undergraduates during a ten-week summer research
period.  We redetected all of the new and previously known pulsars
found in the first processing run conducted at McGill University in
the period range to which we were sensitive ($P \approxgt 20$
ms). This demonstrates the feasibility of this processing approach.
We conclude that the computer cluster and method of analysis described
here can serve as a useful model for scientific computation in the
small college environment in which there are limited resources.  We
hope to modify the existing cluster software for use in other
computationally intensive data processing or modeling projects at
Haverford. The DVD archive of the raw survey data that we have
constructed is reliable, portable, inexpensive, and easily read on any
standard modern machine.  The archive data are available for future
extraction and reprocessing by members of the pulsar community.  We
conclude that for pulsar surveys in which individual files can be
organized not to exceed 4.7 GB (such as the survey described here),
DVD is an attractive alternative format to DLT for data storage.

For the continued development of the cluster, we may seek to revise
existing software for improved performance and reliability, perhaps
though decentralized error checking and recovery and through analysis
of cluster timing results with the intent of finding and removing
bottlenecks.  Furthermore, the existing signal processing and pulsar
search software which was used could be modified.  One possibility
would be to add an ``out-of-core'' FFT so that analysis could be
performed on larger data sets using these machines with their
significant memory limitations, as well as completely RAM-based
processing, perhaps with a RAM disk.  Preliminary tests have shown a
10\% speedup with the use of a RAM disk.  We plan to continue to add
additional machines to the cluster, which is a simple and easy way to
extend to power of the cluster.  The addition of more workstations,
both in the vicinity of the cluster and over the campus network, will
increase the processing output and allow us to test cluster
scalability.  The possibility exists in the future to re-engineer our
cluster to use a SETI@home-style, large-scale distribution system
\citep{a++02}.  Another area of interest involves exploring the impact
of various scheduling, error detection and recovery policies on bother
performance and dependability.  These might include guided
self-scheduling \citep{pk87}, factoring \citep{hsf92}, and
speculation-based policies \citep{hki98}.

\acknowledgements

We thank Dunc Lorimer for providing key components of the software
used in the analysis (Seek and Sigproc) and for his patience when
dealing with our many questions. We thank Scott Ransom for assistance
and for making available portions of the Presto pulsar data processing
code.  We also thank the Haverford College Computer Science Department
for the use of hardware and personnel and the McGill University pulsar
group for extensive collaborative work on the survey on which this
study was based.  This project was supported in part by the Haverford
College Faculty Research Fund, the Haverford Faculty Support Fund, and
the Keck Northeast Astronomy Consortium.

%\clearpage
%
%\begin{figure} 
%\centerline{\psfig{figure=aitoff_pulsar.ps,angle=90,width=7in}}
%\caption{Galactic coordinates plot of the locations of the 56
%unidentified EGRET gamma-ray error boxes surveyed (open circles) and
%the 1300 known pulsars from the public pulsar catalog (solid
%circles). The dashed lines correspond to Galactic latitude $\pm
%5^{\circ}$, the latitude limits of the Parkes Multibeam Pulsar Survey
%\citep{mlc+01}, which had a comparable sensitivity to this survey.
%All of the EGRET targets surveyed lie outside this region.}
%\label{fig-1}
%\end{figure}

\clearpage

\clearpage 

\clearpage

\clearpage 

\begin{deluxetable}{lccccccc}
%\tabletypesize{\scriptsize}
\tablecaption{Cluster Data Processing Times.\label{tbl-1}}
\tablewidth{0pt}
\tablehead{
\colhead{\normalsize Function} 
& {\normalsize client0}
& {\normalsize client1}
& {\normalsize client2}
& {\normalsize client3}
& {\normalsize client4}
& {\normalsize client5\tablenotemark{a}}
& {\normalsize client6}
} 
\startdata
\normalsize download (sec)              & \normalsize $44.5$         & \normalsize $40.6$     & \normalsize $33.2$     & \normalsize $31.4$     & \normalsize $33.9$     & \normalsize $36.7$       & \normalsize $35.9$     \\
\normalsize decimate (sec)              & \normalsize $281.7$        & \normalsize $282.4$    & \normalsize $282.9$    & \normalsize $298.2$   & \normalsize $284.7$    & \normalsize $66.4$       & \normalsize $295.8$    \\ 
\normalsize sc\_td (sec)                & \normalsize $514.9$        & \normalsize $513.8$    & \normalsize $512.9$    & \normalsize $545.9$    & \normalsize $514.8$    & \normalsize $154.7$      & \normalsize $539.3$    \\
\normalsize filterbank (sec)            & \normalsize $55.7$         & \normalsize $49.8$     & \normalsize $41.6$     & \normalsize $53.6$    & \normalsize $55.5$     & \normalsize $19.5$       & \normalsize $54.4$     \\ 
\normalsize hunt trial (sec)\tablenotemark{b} & \normalsize $11.3$   & \normalsize $11.4$     & \normalsize $11.4$     & \normalsize $12.1$    & \normalsize $11.4$     & \normalsize $3.2$        & \normalsize $12.1$     \\ 
\normalsize best (sec)                  & \normalsize $12.3$         & \normalsize $12.9$     & \normalsize $12.7$     & \normalsize $13.1$     & \normalsize $12.6$     & \normalsize $3.6$        & \normalsize $13.3$     \\ \\ 
\normalsize total (min)\tablenotemark{c} & \normalsize $99.9$        & \normalsize $100.3$    & \normalsize $100.4$    & \normalsize $106.7$    & \normalsize $100.8$    & \normalsize $28.3$       & \normalsize $106.5$    \\
\enddata

\tablecomments{Average times for completion of computational functions
in the processing of a single beam of the aggregated pulsar data. See
also Figure \ref{fig-3} for an explanation and graphical
representation of the functions.}

\tablenotetext{a}{2.5-GHz Pentium 4 processor, which also served as a
secondary data server. All other machines were 400-MHz Intel
Celerons.}

\tablenotetext{b}{Average time for completion of a single trial
dedispersion and subsequent FFT. A total of 450 DM trials were
separately searched in hunt. The average time to complete all DM
trials and FFTs for a single beam using hunt is therefore found by
multiplying the listed number by 450.}

\tablenotetext{c}{Average time to fully process a single aggregated
beam (includes all 450 hunt trials).}

\end{deluxetable}

 %TABLE 1: cluster timing info 

\clearpage

\begin{deluxetable}{lccc}
%\tabletypesize{\scriptsize}
\tablecaption{Cost Comparison for Archiving the EGRET Survey 
on DVD vs. DLT-IV Tape.\label{tbl-2}}
\tablewidth{0pt}
\tablehead{
\colhead{Storage Medium} 
& {DVD} 
& {DLT-IV}
& {DVD/DLT Ratio}
} 
\startdata
Number of media used for storage                      & 233 discs & 22 tapes & 10.59 \\
 & & & \\
Cost per medium unit                               & \$1.36 & \$65 & 0.02 \\
 & & & \\
Media writer/reader cost                           & \$400 & \$2300 & 0.17 \\
%Media reader cost                                 & \$0   & \$2300 & \\
 & & & \\
Other costs                                        & \$45\tablenotemark{a} & \$0 & \\
 & & & \\
Maximum storage capacity per                       & 4.7 GB & 35 GB & 0.13 \\
medium unit & & & \\
 & & & \\
Fraction of storage capacity used                  & 0.53 & 0.75 & 0.70 \\
per medium unit & & & \\
 & & & \\
Archive cost per GB (assuming                      & \$0.29/GB & \$1.86/GB & 0.16 \\
maximum storage capacity) & & & \\
 & & & \\
Actual archive cost per GB                         & \$0.55/GB & \$2.48/GB & 0.22 \\ 
 & & & \\
Total archive cost                                 & \$760 & \$3730 & 0.20 \\
\enddata

\tablecomments{All prices from June 2003.}

\tablenotetext{a}{DVD storage case, which holds 264 discs.} 

\end{deluxetable}

 %TABLE 2: DVD/DLT cost comparison.

\end{document}